\documentclass[aps,pre,reprint,superscriptaddress]{revtex4-1}

\usepackage{graphicx}
\usepackage{amssymb,amsmath}
\usepackage{textcomp} 
\usepackage[breaklinks]{hyperref}
\usepackage[usenames,dvipsnames]{color}

\begin{document}
\author{Thomas A Caswell}
\affiliation{The James Franck Institute and Department of Physics,\\ The University of Chicago, Chicago, Illinois 60637, USA}

\author{Zexin Zhang}
\affiliation{Center for Soft Condensed Matter Physics and Interdisciplinary Research, \\ Soochow University, Suzhou 215006, China, P. R.}


\author{Margaret L Gardel}
\affiliation{The James Franck Institute and Department of Physics,\\ The University of Chicago, Chicago, Illinois 60637, USA}

\author{Sidney R Nagel}
\affiliation{The James Franck Institute and Department of Physics,\\ The University of Chicago, Chicago, Illinois 60637, USA}

\title{Observation and Characterization of the
  Vestige of the Jamming Transition in a Thermal 3D System}
\date{\today}

\begin{abstract}
We study the dependence of  the pair-correlation function, $g(r)$, and particle mobility on packing fraction in a dense three-dimensional packing of soft colloids made of poly N-isopropyl acrylamide (pNIPAM), a thermo-sensitive gel. We find that $g(r)$ for our samples is qualitatively like that of a liquid at all packing fractions.  There is a peak in $g_{1}$, the height of the first peak of $g(r)$, as a function of packing fraction.  This peak is identified as the thermal remnant of the $T=0$ divergence found at the jamming transition in simulations of soft frictionless spheres at zero-temperature.   Near where there is a peak in $g_{1}$ the particles become arrested on the time scale of the experiment.

\end{abstract}

\maketitle

\section{Introduction}
\label{sec-1}
From hard-packed granular road beds to glass look-out ledges in skyscrapers, amorphous materials are used as solid support structures. Yet we do not understand the origin of rigidity in those materials. In the case of crystalline solids we understand the onset of rigidity as being a consequence of breaking translational symmetries during the fluid-to-crystal transition~\cite{chakin}.  In contrast, no obvious symmetries are broken in the transition of a fluid to an amorphous solid. Is there any structural signature, however subtle, that can be identified with the onset of rigidity?  Previous work has searched for such a signature in the pair-correlation function, $g(r)$.  Different criteria have been proposed.  These include the splitting of the second peak of $g(r)$ into two sub-peaks~\cite{AlfonsvanBlaaderen11171995},the ratio of the first minimum to the first maximum of $g(r)$~\cite{PhysRevLett.41.1244}, and changes to the contact-force distribution which is related to $g(r)$ at small $r$~\cite{corwin2005structural}.

Jamming is a way of constructing rigid amorphous structures~\cite{liu1998jnj}. Recent simulations of jamming at zero temperature~\cite{PhysRevE.68.011306} have suggested a different structural signature for rigidity onset.  The rigidification of random packings of finite-range soft repulsive spheres at temperature $T=0$, is controlled by the packing fraction, $\phi$.  Such packings have a sharp jamming transition at a critical packing fraction, $\phi_{c}$, where the particles first unavoidably make contact.  Below $\phi_{c}$ no particles overlap, whereas above $\phi_{c}$ particles must overlap with their nearest neighbors and the system can support stress.  This onset of rigidity is a purely geometric effect.  At $\phi_{c}$ all nearest-neighbor pairs are separated by precisely one particle diameter. This leads to a $\delta$-function in $g(r)$ at its first peak.  This divergence is a signature of the athermal (i.e., $T=0$) jamming transition and varying $\phi$ above or below $\phi_{c}$ suppresses the divergence~\cite{PhysRevE.68.011306}.

The relevance of jamming to systems at $T>0$ as well as its relation to dynamical arrest and the glass transition have not been clear. Dense suspensions of hard colloids have been thoroughly studied in the context of the colloidal glass transition and super-cooled fluids~\cite{ISI:000234227800002,0953-8984-19-20-205131,PeterSchall12212007,conrad:265701,0953-8984-15-1-351,weeks2002subdiffusion,PhysRevLett.89.095704,weeks00:_three_dimen_direc_imagin_struc,kegel2000direct,AlfonsvanBlaaderen11171995,lynch-2008-78,0953-8984-19-20-205131,conrad:265701,cianci2006correlations,0953-8984-15-1-349,weeks00:_three_dimen_direc_imagin_struc}. However, hard colloids are ill suited to study many aspects of jamming since they cannot overlap so can only access configurations on the fluid side of jamming, that is $\phi<\phi_{c}$. In this paper, we study as a function of $\phi$ a three-dimensional packing of soft colloids made from a thermo-sensitive hydrogel. Our particles are small enough to undergo Brownian motion and thus can be considered to be at finite temperature.  To search for a structural signature of the jamming transition, we use optical confocal microscopy and particle tracking techniques to determine the positions and displacements of particles in order to calculate the pair-correlation function, $g(r)$, and the particle mobility.

In order to investigate the structure of our samples as a function of packing fraction we control $\phi$ by varying the particle diameter at a fixed number density. As we vary $\phi$, the height of the first peak of $g(r)$, $g_{1}$, varies in a non-monotonic fashion and has a maximum at $\phi = \phi^{*}$, near where there is also a dramatic reduction in the particle mobility. This maximum is a vestige of the zero-temperature jamming transition.  The value of $\phi^{*}$ is greater than $\phi_{c}$, the packing fraction where jamming would occur at $T=0$.  These observations are qualitatively consistent with recent simulations~\cite{1112.2429} and two-dimensional experiments using bi-disperse soft colloids~\cite{Zhang2009}.  

The fact that the first peak in $g(r)$ is largest at $\phi^{*}$ suggests that the sample at that $\phi$ may have more long range order than at other packing fractions.  This can be checked by looking at the subsequent peaks of $g(r)$ which contain information about medium- and long- range order. Close to and above $\phi^{*}$ we observe that $g(r)$ has at least 14 evenly spaced peaks which decay in height as $r$ increases. In this regime the subsequent peaks do not vary appreciably with $\phi$. However, lowering $\phi$ below $\phi^{*}$ strongly increases the damping of the higher-order peaks.  We see no evidence of a split second peak in $g(r)$ as is seen in hard-colloidal systems at the colloidal glass transition~\cite{Jenkins200865}.

We also compare the structure and mobility of our samples as a function of packing fraction. We find a pair of systems on opposite sides of $\phi^{*}$ with an order of magnitude difference in the particle mobility where the pair-correlation functions are experimentally indistinguishable.  This suggests that the dynamics cannot be predicted simply from a knowledge of the structure.

\section{Methods/materials}
\label{sec-2}

 \begin{figure}
 \centering
 \includegraphics[width=8.5cm]{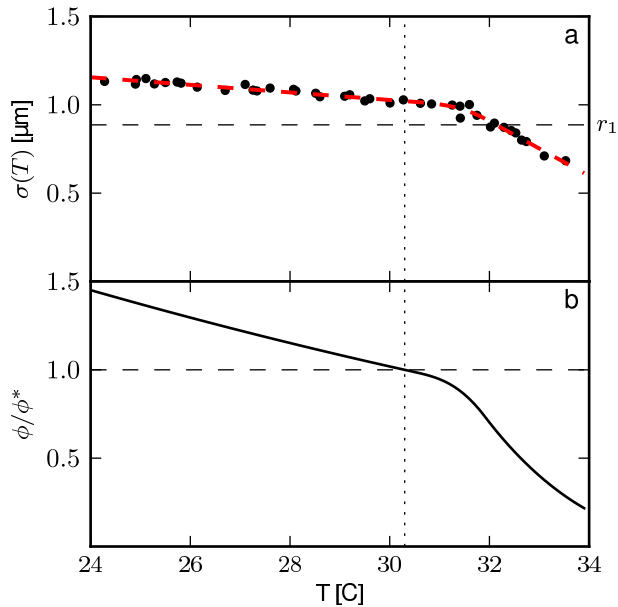}
 \caption{\label{fig:p_size}a) The hydrodynamic diameter of the pNIPAM colloids used in the experiments.  Dashed line is an empirical fit to the data (solid circles).  The horizontal line is the average nearest neighbor spacing for a given number density, $n$.  b) Relative packing fraction.  Vertical line in both graphs is the temperature where $\phi = \phi^{*}$}
\end{figure}

 \begin{figure}
 \centering
 \includegraphics{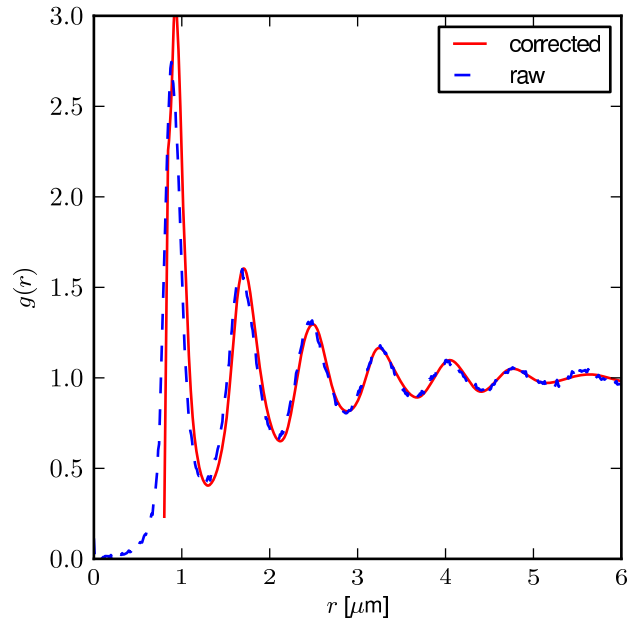}
 \caption{\label{fig:corr} The corrected and raw $g(r)$ curves for a    sample at $\phi = \phi^{*}$.}
 \end{figure}

We synthesize poly N-isopropyl acrylamide  (pNIPAM) colloids using surfactant-free emulsion polymerization~\cite{Pelton20001,saunders1999microgel}.  The size, stiffness, and temperature dependence of the colloids is highly sensitive to the details of the synthesis and each batch must be calibrated independently. The hydrodynamic particle diameter, $\sigma$, depends on the environmental temperature, $T$.  We measure $\sigma(T)$ by observing the diffusion of the colloids in a dilute sample.  As shown in Figure \ref{fig:p_size}a, over the range $24$\textdegree C$<T<34$\textdegree C, the diameter varies by nearly a factor of two.  The data for $\sigma(T)$ is empirically fit by two linear segments connected by a cubic spline.  Above $35$\textdegree C, the colloids collapse to a constant size\cite{Pelton20001}.  The sample temperature, measured to a precision of 0.1\textdegree C, is controlled with a Bioptechs objective heater thermally coupled to the sample via the objective immersion fluid.

In this paper, we estimate the packing fraction, $\phi$, of dense systems from the particle diameter, $\sigma$ and the number density, $n$,
\begin{equation}
  \label{eq:2}
  \phi(T) = \frac{\pi}{6} n \sigma(T)^{3}.
\end{equation}
Over a 10\textdegree C temperature range, $\phi$ varies by a factor of 7 as shown in Figure\ref{fig:p_size}b. We take data in the range between $T=27$\textdegree C to $32$\textdegree C.  Over this range,the sample transforms from a liquid state, where the particles are diffusive, to a fully arrested packing.  The primary effect of changing the environmental temperature is to change the size of the colloids (and hence $\phi$) rather than the thermal energy.

Sample chambers are made from microscope slides and cover slips and sealed with epoxy (Norland 61).  The chambers are small, $0.15$mm $\times$ $5$mm $\times$ $3$mm, to minimize thermal gradients within the sample.  To prepare dense samples, sparse suspensions of colloids are centrifuged to sediment the colloids.  The suspension is then heated to 35\textdegree C to de-swell the particles. The sediment is then pipetted into the sample chamber which is quickly sealed with epoxy so that the total volume, particle number, and number density, $n$, are fixed.  After enclosure in the chamber, the samples are prepared for observation by first heating to $35$\textdegree C to turn the system into a fluid.  The sample is then cooled to $4$\textdegree C in less than 5s by placing the sample chamber on $4$\textdegree C metal surface.  This quench protocol consistently generates arrested amorphous configurations.

Particle dynamics and $g(r)$ are extracted from data sets taken at different values of $\phi$.  Between subsequent measurements, the sample is melted and re-quenched.  The waiting time from the quench to observation is approximately 900 seconds and is controlled to minimize possible complications due to aging\cite{mazoyer:011501,0953-8984-15-1-349}.  In order to sample the $\phi$ dependence of $g(r)$ in finer increments the packing fraction was also swept continuously.  For increasing $\phi$ ramps, a fluidized sample is placed on a pre-heated objective. The objective heater is then turned off and the sample is periodically imaged as it cools slowly to ambient temperature.

We image a two-dimensional slab far from the boundaries of a three-dimensional sample using a Yokogawa CSU-XI confocal head and a Nikon 60x water immersion objective.  Images are acquired at frame rates between $0.3$Hz and $10$Hz using a 14 bit-depth Roper Coolsnaps HQ camera.  The field of view is $150$\textmu m $\times$ $111$\textmu m and contains approximately $20,000$ particles. The accuracy of feature identification in the $x$-$y$ plane is approximately $15$nm which corresponds to approximately $0.1$ of a pixel.

Due to the confocal slice having a finite thickness, $w$, particles both above and below the focal plane are imaged as being in the plane.  This introduces an uncertainty, $w/2 \approx 400$nm, in the $z$ position of the particle.  As a result the measured distance between two particles, which is the distance projected onto the $x-y$ plane, will be less than the true distance.  The difference between the measured and true distance becomes small when $r\gg w$ but can be significant at small $r$. This introduces a systematic error in $g(r)$ which can be corrected~\cite{JPSJ.78.065004}  
\begin{equation}
\label{eq:gofr_corr}
\begin{split}
  g(r) \approx& g_{\rm{raw}}\left[r\left(1 - \frac{1}{12} \frac{w^{2}}{r^{2}} + \frac{1}{720} \frac{w^{4}}{r^{4}}\right)\right]\\
  &\quad + \frac{7}{1440} \frac{w^{4}}{r^{4}} r^{2} \left . \frac{{\rm d}^{2} g_{\rm{raw}}}{{\rm d}r^{2}}\right |_{r}
\end{split}
\end{equation} 
where  $g_{\rm{raw}}$ is the pair-correlation function measured prior to taking into account the effects of projection and $r>w$. We have used simulations to check the validity of this approximation. 

The correction shifts peaks outward, increases peak heights and decreases the valleys of $g(r)$. It primarily affects the first peak of $g(r)$. This shown in Fig.~\ref{fig:corr}. The position of the first peak is shifted by $4\%$ and its height is enhanced by $10\%$; the effect on higher order peaks is significantly smaller. At large $r$ the primary source of the noise in $g(r)$ is from under-sampling configuration space. The projection is present in all of our data sets, thus will not effect relative measurements as a function of $\phi$.

The data is processed using a locally developed C++ implementation of the Crocker-Grier feature identification and tracking algorithm~\cite{crocker95_meth_dig} incorporating an existing implementation of the identification routine~\cite{lu2007_TARC}.  The tracking and correlations software is locally developed and is available at \url{https://github.com/tacaswell/tracking}.  Our tracking and correlation code is an order of magnitude faster than equivalent code written in an interpreted language.  The implementation of spatial correlation functions scales as $O(n \times r_{\rm{max}}^{d})$ where $d$ is the dimension of the correlation and $n$ is the total number of particles.

\section{Results and Discussion}

 \begin{figure}
 \centering
 \includegraphics[width=8.5cm]{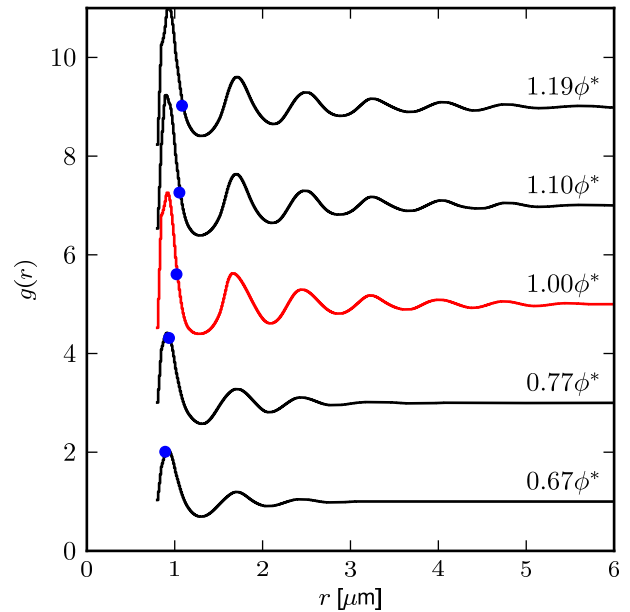}
 \caption{\label{fig:gofr_all} The pair-correlation functions, $g(r)$, for a range of packing fractions, $\phi$. The data has been corrected for the projection effects discussed in the text. The bottom curve is for $\phi < \phi_{c}$, for all other curves $\phi>\phi_{c}$.  The red curve is for $\phi = \phi^{*}$.  The position of the first peak remains fixed even though the nominal particle diameter, indicated by the dots at each packing fraction, varies. }
 \end{figure}

We will first discuss some of the general features of $g(r)$ and then in the subsections below focus on specific aspects of the data. 

Figure \ref{fig:gofr_all} shows the pair-correlation function at different the packing fractions, $\phi$.  This data has been corrected for the projection issues as discussed in Section \ref{sec-2}.  In each case, there is a large first peak at $r_1 \approx 0.9 \mu m$.  At the lowest $\phi$ shown, only the first three peaks are large enough to be easily discerned.  As $\phi$ is increased, peaks at larger $r$ grow and become clearly visible.  

The position of the first peak, $r_1$, does not shift appreciably even though the packing fraction varies by a factor of 1.7.  To emphasize this point, the nominal particle diameter, $\sigma$, is marked on each curve in Figure \ref{fig:gofr_all}.  At all $\phi > \phi_{c}$, $\sigma > r_{1}$, indicating that the particles strongly interact with and deform their neighbors.  As we vary $\sigma$, or equivalently $\phi$, the height of $g_{1}$ varies but the peak postions do not move in location.  This can be understood because the average interparticle spacing in this densely packed system is set by the number density, $n$ which is held constant by the constraint that the number of particles in the rigid sample cell is not allowed to vary.
   
To determine the absolute packing fraction for our system, we first determine the number density of our system since, at $\phi_{c}$ and $T=0$, $r_{1} = \sigma$.   
Substituting into equation (\ref{eq:2}) we find $n = \frac{6\phi_{c}}{\pi r_{1}^{3}}$ and
\begin{equation}
  \label{eq:phirsigma}
  \phi = \phi_{c} \left(\frac{\sigma}{r_{1}}\right)^3
\end{equation} 
We use $\phi_{c} = 0.64$, which is the density of random close packing as measured in simulations and hard sphere experiments~\cite{PhysRevE.68.011306,scott62:_radial_distr_random_close_packin_equal_spher}. Our samples range in packing fractions from $\phi =0.55$ to $\phi =1.17$.

\subsection{Behavior of First peak of $g(r)$: Vestige of the jamming transition}

 \begin{figure}
 \centering
 \includegraphics[width=8.5cm]{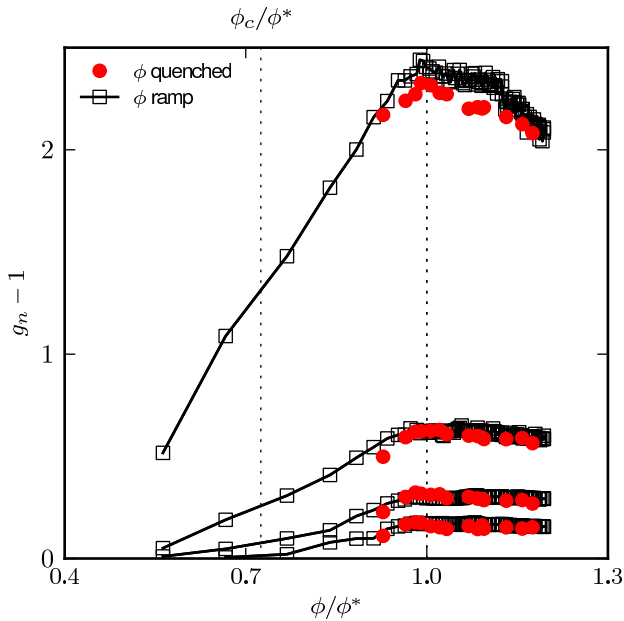}
 \caption{\label{fig:continous_temp}Peak heights versus normalized packing fraction. $g_{m}-1$ is plotted versus $\phi/\phi^{*}$ for the first four peaks, $m= 1,...,4$.  $\square$ are extracted from data where the volume fraction is continuously ramped to higher values. {\color{red}$\bullet$} is extracted from data quenched to $\phi$.  At all volume fractions the vertical error bars are on the order of the marker size.  For $\phi > \phi^{*}$ the data points are tightly spaced. }
 \end{figure}

In figure \ref{fig:continous_temp}, we plot the height of the $m^{\rm{th}}$ peak, $g_{m}$, of the first four peaks in $g(r)$ versus $\phi$. We measured $g_{m}$ in two protocols. One is to quench the sample rapidly to a desired packing fraction and the other is slowly ramp $\phi$. The two protocols yield nearly identical results showing that the data is not significantly affected by aging or by transient effects. 

The first peak, $g_{1}$, has non-monotonic behavior. Starting at small $\phi$, $g_{1}$ grows with $\phi$ until it reaches a maximum value at $\phi \equiv \phi^{*}$ and then decreases as $\phi$ is increased further. The other peaks, $m=2,3,4$, appear to reach a plateau above $\phi^{*}$.  Within the error bars of our measurement, they show no peak. The behavior in $g_{1}$ is consistent with experiments~\cite{Zhang2009} and simulations~\cite{0911.1576v1,1112.2429} on thermal two-dimensional bi-disperse systems. As in those cases, we interpret the peak in $g_{1}$ as a vestige of the $T=0$ jamming transition.

The behavior of $g_{1}$ is due to the interplay between particle overlap caused by thermal motion and particle overlap due to geometric frustration, as described in~\cite{Zhang2009}. This leads to a peak in $g_{1}$ as a function of $\phi$ as observed.  The maximum in $g_{1}$ is necessarily shifted to $\phi>\phi_{c}$ by the thermal motion. In our three-dimensional samples, $\phi^{*}$ is shifted above $\phi_{c}$ by a much greater amount than was reported in the two-dimensional studies.  

\subsection{Structure: long-range correlations}

 \begin{figure}
 \centering
 \includegraphics{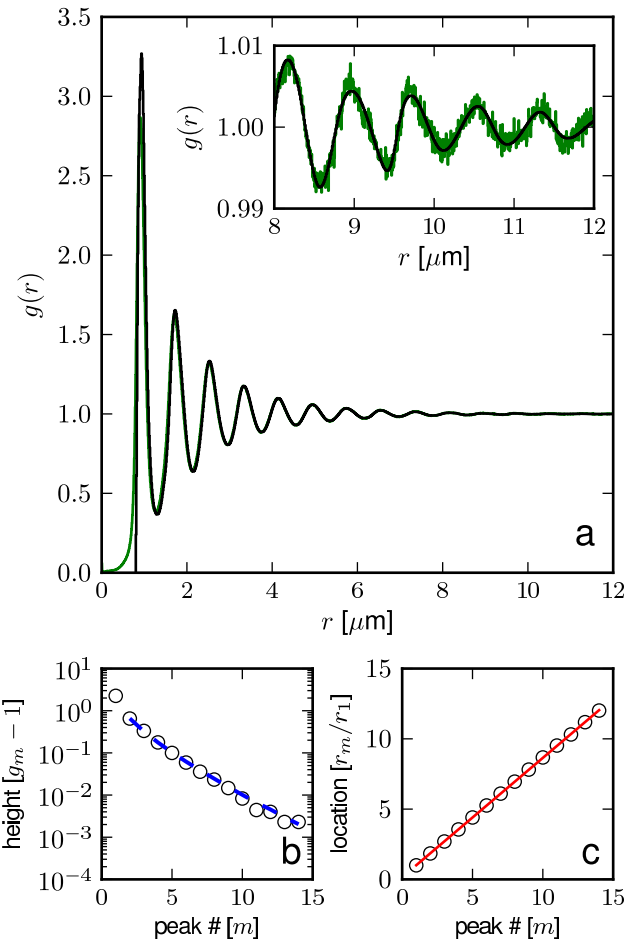}
 \caption{\label{fig:gofr} a) The pair-correlation function for $\phi = 0.98\phi^{*}$.  The raw $g(r)$ is shown in green and the corrected $g(r)$, eq (\ref{eq:gofr_corr}), is shown in black. The inset is a zoom of $8$\textmu m$<r<12$\textmu m to show the detail of the higher order peaks.  b) Peak height versus peak number $m$. Peak height are fit to the decay form of eq (\ref{eq:pk_hta}), shown in blue.  c) Peak location versus $m$. The peak locations are evenly spaced and can be fit to straight line, eq. (\ref{eq:rn}), with a slope of $0.85$.}
 \end{figure}

The data in Figure \ref{fig:gofr_all}  shows no sign of a split second peak.  This is in marked contrast to studies of hard spheres~\cite{AlfonsvanBlaaderen11171995} and simulations at $T=0$~\cite{PhysRevE.68.011306,silbert:041304}.  In those systems, the second peak splits into two sub-peaks as the packing fraction approaches $\phi_{c}$.  This splitting has been used as a possible signature to identify colloidal glasses~\cite{Jenkins200865}.  However, we do not see the second peak split at any $\phi$ in our amorphous packings.  

At large $\phi$, many peaks, corresponding to higher-order coordination shells, are visible.  The height of these peaks decays at large $r$.  Remarkably, as shown in  Figure \ref{fig:gofr}, near $\phi^{*}$ we can identify at least 14 peaks in $g(r)$, the smallest of which are fluctuations of less than $1\%$ from uniform density as seen in the inset.  We are unaware of any other experiment that identifies as many peaks in an amorphous sample. 

To study the form of $g(r)$ we plot the peak height, $g_{m}$, and location, $r_{m}$, against peak number $m$ in Figure \ref{fig:gofr} b) and c) respectively.  We can fit the decay envelope to the Percus-Yevic asymptotic form,
\begin{equation}
  \label{eq:pk_hta}
  g_{m} - 1 = C \frac{\exp\left(\frac{r_{m}}{\xi}\right)}{r_{m}},
\end{equation}
where $r_{m}$ is the location of the $m^{\rm{th}}$ peak and $C$ and $\xi$ are fitting parameters which depend on $\phi$~\cite{perry:1827}.  At $\phi^{*}$ $C = 2.31$ and $\xi = (-2.6\pm.1) r_{1}$, which is a longer correlation length than seen in hard sphere simulations and experiments~\cite{PhysRevLett.95.090604,1003.3965}.

Beyond the first peak, the inter-peak spacing is very uniform: the location of the $m^{\rm{th}}$ peak, $r_{m}$, is accurately given by
\begin{equation}
  \label{eq:rn}
  r_{m}/r_{1} = 1+(0.85\pm0.02)(m-1)
\end{equation}
as shown in Figure \ref{fig:gofr} c. This is also true for the higher-order peaks at other values of $\phi$ where they are visible. The peak spacing is in quantitative agreement with what has been observed in other systems: hard-sphere colloids~\cite{1003.3965}, ball bearings~\cite{scott62:_radial_distr_random_close_packin_equal_spher}, simulations at $\phi_{c}$~\cite{PhysRevLett.95.090604} and with experimental measurements of liquid noble gases~\cite{scott62:_radial_distr_random_close_packin_equal_spher}, despite the drastically different inter-particle potentials and temperatures.  This suggest a fundamental geometric origin for this peak spacing.

These two results taken together suggest that beyond the first peak, $g(r)$ can be approximated by a damped sinusoid:  \begin{equation}
\label{eq:dmp_sin}
g(r) = 1 + \frac{C}{r}\exp\left(\frac{r}{\xi}\right) \cos\left(2 \pi \frac{r-r_{1}}{0.85 r_{1}}\right)
\end{equation}
This form is roughly consistent with a wide range of analytic~\cite{matteoli:4672}, simulation~\cite{PhysRevLett.95.090604,heyes:204506}, and experimental~\cite{1003.3965} results and indicative of a dense fluid structure.

\subsection{Dynamics: Slowing at $\phi^{*}$}
 \begin{figure}
 \centering
 \includegraphics[width=8.5cm]{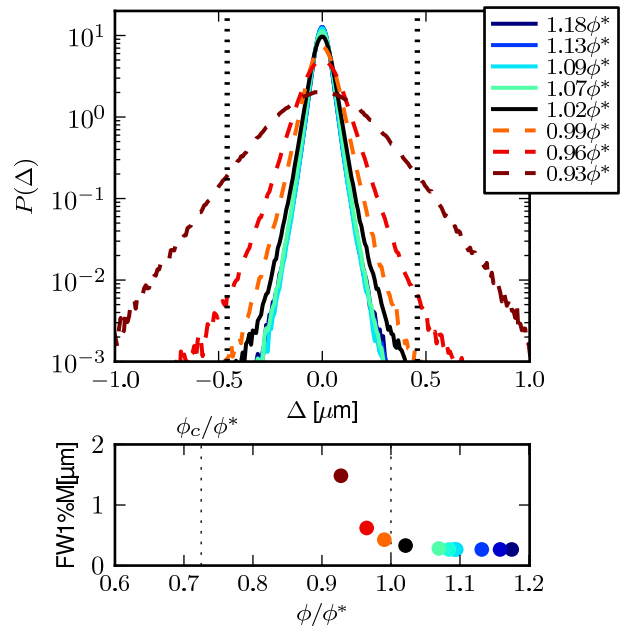}

 \caption{\label{fig:van_Hove_shape}a) van Hove correlation functions for $\tau = 1.5s$.  Curves for $\phi< \phi^{*}$ are dashed (- -) and curves for $\phi> \phi^{*}$ are solid (--).  The curves for $\phi>1.09\phi^{*}$ cannot be distinguished on this plot because they lie under the curve at $1.07\phi^{*}$. The vertical dotted lines indicate the average particle radius. b) The full width at $1\%$ max as a function of $\phi$.}
 \end{figure}

 \begin{figure}
 \centering
 \includegraphics[width=8.5cm]{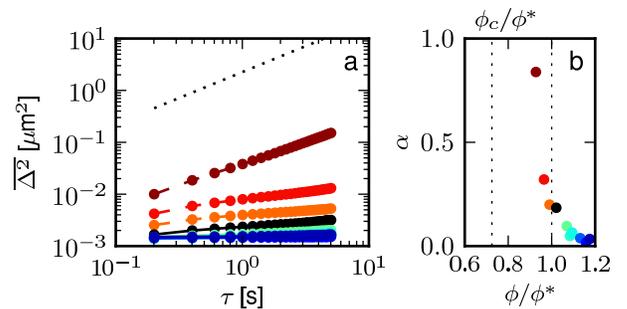}
 \caption{\label{fig:van_Hove_msd}Mean squared particle displacement as a function of $\tau$ for different values of $\phi$.  The dotted line indicates the diffusion in a dilute sample.  The symbols and colors match those in Figure \ref{fig:van_Hove_shape}.}
 \end{figure}

We now ask whether packing fraction, $\phi^{*}$, where $g_{1}$ has a peak, is associated with any change in the dynamics in the sample.  To quantify the particle mobility we use the van Hove correlation function,
\begin{equation}
\label{eq:vanHove}
P_{x}(\Delta,\tau) = \frac{1}{N}\left\langle \sum_{i} \delta\left(x_{i}(t) - x_{i}(t+\tau) - \Delta\right) \right\rangle,
\end{equation}
where $x_{i}(t)$ is the $x$ component of the $i$th particle location at time $t$, $\langle\rangle$ is an average over all starting times, and $N$ is a normalization constant such that $\int \rm{d}\Delta\, P_{x}(\Delta,\tau) = 1$.  Physically $P_{x}(\Delta,\tau)$ is the probability of a particle moving a distance $\Delta$ in the $x$ direction in a time $\tau$.  We can similarly define $P_{y}(\Delta,\tau)$ for motion in the $y$ direction.  The mobility is isotropic in the $x$ and $y$ directions, which allows us to average these distributions to improve statistics at large $\Delta$.  We call the averaged distribution $P(\Delta,\tau)$. We can compute the three-dimensional mean-squared displacement as the second moment of $P(\Delta,\tau)$,
\begin{equation}
  \label{eq:msd}
  \overline{\Delta^{2}}(\tau) = 3\int \rm{d}\Delta\, \Delta^{2}P(\Delta,\tau).
\end{equation}
For a diffusive system $P(\Delta,\tau)$ is a Gaussian centered on $\Delta = 0$ and $\overline{\Delta^{2}}(\tau) \propto \tau^{\alpha}$ with $\alpha = 1$.

There is a qualitative change in particle mobility as the packing fraction approaches $\phi^{*}$.  This change is evident is the shape of $P(\Delta,\tau=1.5 s)$ shown in Figure \ref{fig:van_Hove_shape}. Curves for $\phi<\phi^{*}$ and $\phi>\phi^{*}$ are plotted as dashed and solid lines respectively. The vertical dashed lines are separated by $r_{1}$ and indicate at which point particles have moved half the distance to their neighbors, $r_{1}/2$, in $1.5 s$.  

At the lowest $\phi$ plotted, $0.93\phi^{*}$, the distribution is nearly Gaussian and a significant number of particles move more than $r_{1}/2$, indicating substantial rearrangement of the packing on the $1.5 s$ time scale.  We cannot take dynamical measurements below $0.93 \phi^{*}$, because, over the course of a few minutes, crystalline regions begin to nucleate.  As $\phi$ is increased, $P(\Delta)$ narrows.  Here we only report dynamic measurements where the sample is stable over the entire course of the experiment.  Static measurements can be made from a single snapshot and can thus be acquired from meta-stable configurations.  Above $1.07\phi^{*}$ the curves are nearly indistinguishable and lie on top of each other.  In this dense state, the particles are essentially arrested on the time scales probed.  Even at these large values of $\phi$, the error in finding particle positions is much smaller than the width of the distributions measured.

To quantify the change in particle mobility, we compute $\overline{\Delta^{2}}(\tau)$, plotted in Figure \ref{fig:van_Hove_msd}a. For comparison, the expected $\overline{\Delta^{2}}(\tau)$ for a dilute sample is shown as the dotted line.  At the lowest $\phi$ the particles are nearly diffusive, but with the diffusion constant significantly reduced from the dilute limit.  The narrowing of $P(\Delta,\tau)$ is reflected in the vertical shift of the curves and is accompanied by a suppression of the slope.  If we assume that $\overline{\Delta^{2}}(\tau) \propto \tau^{\alpha}$, then the by fitting the slope we can extract $\alpha(\phi)$, shown in Figure \ref{fig:van_Hove_msd}b.  There is clearly a dynamical transition in the neighborhood of $\phi^{*}$, but it does not have a sharp signature. Surprisingly, there is a range of $\phi$ where the particles are in contact with their neighbors, i.e., $\phi>\phi_{c}$, but the system is nevertheless nearly diffusive.  It is tempting to associate the dynamical arrest at $\phi^{*}$ with the vestige of the jamming transition.  However this is problematic because the peak in $g_{1}$ is an unambiguous geometric signature and does not display aging effects, whereas the dynamical signature is not sharp and depends on the experimental timescale used to perform the measurements.

 \begin{figure}
 \centering
 \includegraphics[width=8.5cm]{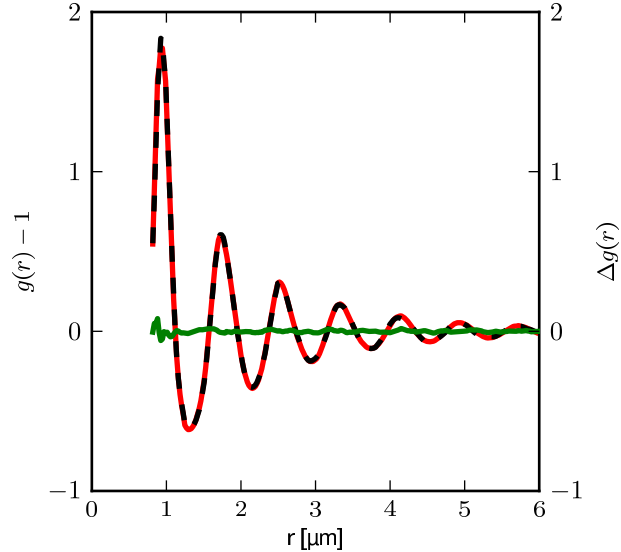}
 \caption{\label{fig:gofr_diff} The pair-correlation function,  $g(r)$, at $0.96\phi^{*}$ and $1.06\phi^{*}$ (on left axis).  The difference between these curves, $\Delta g$, (right axis) is smaller than the experimental resolution despite the drastic difference in particle mobility.}
\end{figure}

It has been argued that the particle mobility can be derived from the pair-correlation function~\cite{1742-5468-2005-05-P05013}. However, if we compare $g(r)$ and $\alpha$ at $0.96\phi^{*}$ and $1.05\phi^{*}$, we see that despite an order of magnitude difference in particle mobility, the structure is experimentally indistinguishable.  This is in agreement with recent work that showed that, by using different potentials, configurations with very similar $g(r)$ can have very different dynamics~\cite{berthier:214503}.  These results are potentially problematic for mode coupling theory as we have experimentally shown that the same static structure can give rise to very different dynamics~\cite{Berthier2010}.

\section{Conclusions}

We have demonstrated that there is a peak in $g_{1}$ as a function of $\phi$ in a three-dimensional packing of soft pNIPAM colloids undergoing Brownian motion.  This is a vestige of the $T=0$ jamming transition that survives at finite temperature and is consistent with previous experiments in two-dimensional systems~\cite{Zhang2009}.  However, in contrast with the results in two-dimensions, the packing fraction, $\phi^{*}$, where $g_{1}$ has a peak is significantly higher than $\phi_{c}$, the packing fraction where particles would first jam at zero temperature.  This indicates that, although the system is very far from the jamming point, aspects of the jamming transition are still observed in the structure of the sample.  
We have also shown that our soft colloid fluid will become dynamically arrested when the packing fraction is increased above $\phi_{c}$.  This is an example of a pressure induced glass transition~\cite{PhysRevE.73.040501,liu1998jnj,doi:10.1021/j100488a007,PhysRevB.72.094204,shumway:1796}.   We observe that the packing fraction where dynamical arrest occurs is close to but not identical with  $\phi^*$.  

We have also observed that the pair-correlation function for this soft-sphere system has features expected of a fluid.  For example, the second peak in $g(r)$ is smooth with no sign of any splitting that is characteristic of hard-sphere systems.  Moreover, near $\phi^{*}$, we observe up to $14$ equally spaced peaks in $g(r)$ whose amplitude decays as a function of distance in a fashion that is also consistent with predictions of liquid structure.  As the packing fraction is decreased below $\phi^{*}$, the damping of the peaks becomes much more dramatic so that near $\phi_{c}$ only three peaks are clearly visible.   We observe that in Fig. 4, $g_{1}$ increases rapidly at low $\phi$ until it reaches a peak, but then decreases more gradually above $\phi^*$.  This asymmetry around the peak can be related to how the contributions to the overlap of particles from thermal motion and from pressure vary with packing fraction.  The role of temperature in broadening the first peak rapidly becomes less important as the packing fraction is increased.

Because the pNIPAM colloids are so soft, it is possible to have very large particle overlap.  This not only changes the structure of $g(r)$ but also allows the system to be diffusive at packing fractions, $\phi$, that are inaccessible to hard-spheres.  To understand thermal systems at high densities, such as molecular glass formers, it is important to understand the dynamics of soft-sphere packings.  Further studies are needed on the dynamics in the neighborhood of $\phi^{*}$, particularly in comparison to the dynamical heterogeneity and correlated motion observed in hard-sphere packings near $\phi_{c}$~\cite{0953-8984-19-20-205131}.  This system can also be used to measure the density of states to determine if the dynamical predictions at $T=0$ extend to $T>0$ in three dimensions.

\section{Acknowlegments}
We are grateful to Justin Burton, Andrea Liu, Peter Lu, and Ning Xu for helpful discussions.  We particularly thank Arjun Yodh for his help and advice on the experiment.  SRN was supported by the U.S. Department of Energy, Office of Basic Energy Sciences, Division of Materials Sciences and Engineering under Award DE-FG02-03ER46088.  TAC and MLG were supported by the University of Chicago NSF Materials Research Science and Engineering Centers DMR-0820054

\bibliographystyle{apsrev4-1}

\end{document}